\documentclass[aps,prl,twocolumn,superscriptaddress,showpacs,showkeys]{revtex4}
\usepackage{graphicx}
\usepackage{amsmath}

\def\NH3{NH$_3$}
\def\ND3{ND$_3$}
\def\Q2{$Q^2$}
\def\ChPT{$\chi$PT}
\def\F1{$F_1$}
\def\F2{$F_2$}
\def\g1{$g_1$}
\def\g1p{$g_1^p$}
\def\g1d{$g_1^d$}
\def\g2{$g_2$}
\def\etal{{\it et al.}}
\def\Cerenkov{\v{C}erenkov}

\begin{document}

\def\deeps{\mbox{D($e,e^\prime p_s$)}}
\def\f2neff{$F_{2n}^{\mbox{\tiny eff}}$}

\title{Moments of the Spin Structure Functions $g_1^p$ and $g_1^d$ 
for $0.05 < Q^2 < 3.0$ GeV$^2$}
\newcommand*{\ANL}{Argonne National Laboratory}
\affiliation{\ANL}
\newcommand*{\ASU}{Arizona State University, Tempe, Arizona 85287-1504}
\affiliation{\ASU}
\newcommand*{\UCLA}{University of California at Los Angeles, Los Angeles, California  90095-1547}
\affiliation{\UCLA}
\newcommand*{\CSU}{California State University, Dominguez Hills, Carson, CA 90747}
\affiliation{\CSU}
\newcommand*{\CMU}{Carnegie Mellon University, Pittsburgh, Pennsylvania 15213}
\affiliation{\CMU}
\newcommand*{\CUA}{Catholic University of America, Washington, D.C. 20064}
\affiliation{\CUA}
\newcommand*{\SACLAY}{CEA-Saclay, Service de Physique Nucl\'eaire, 91191 Gif-sur-Yvette, France}
\affiliation{\SACLAY}
\newcommand*{\CNU}{Christopher Newport University, Newport News, Virginia 23606}
\affiliation{\CNU}
\newcommand*{\UCONN}{University of Connecticut, Storrs, Connecticut 06269}
\affiliation{\UCONN}
\newcommand*{\ECOSSEE}{Edinburgh University, Edinburgh EH9 3JZ, United Kingdom}
\affiliation{\ECOSSEE}
\newcommand*{\FU}{Fairfield University, Fairfield CT 06824}
\affiliation{\FU}
\newcommand*{\FIU}{Florida International University, Miami, Florida 33199}
\affiliation{\FIU}
\newcommand*{\FSU}{Florida State University, Tallahassee, Florida 32306}
\affiliation{\FSU}
\newcommand*{\GWU}{The George Washington University, Washington, DC 20052}
\affiliation{\GWU}
\newcommand*{\ECOSSEG}{University of Glasgow, Glasgow G12 8QQ, United Kingdom}
\affiliation{\ECOSSEG}
\newcommand*{\ISU}{Idaho State University, Pocatello, Idaho 83209}
\affiliation{\ISU}
\newcommand*{\INFNFR}{INFN, Laboratori Nazionali di Frascati, 00044 Frascati, Italy}
\affiliation{\INFNFR}
\newcommand*{\INFNGE}{INFN, Sezione di Genova, 16146 Genova, Italy}
\affiliation{\INFNGE}
\newcommand*{\ORSAY}{Institut de Physique Nucleaire ORSAY, Orsay, France}
\affiliation{\ORSAY}
\newcommand*{\ITEP}{Institute of Theoretical and Experimental Physics, Moscow, 117259, Russia}
\affiliation{\ITEP}
\newcommand*{\JMU}{James Madison University, Harrisonburg, Virginia 22807}
\affiliation{\JMU}
\newcommand*{\KYUNGPOOK}{Kyungpook National University, Daegu 702-701, South Korea}
\affiliation{\KYUNGPOOK}
\newcommand*{\MIT}{Massachusetts Institute of Technology, Cambridge, Massachusetts  02139-4307}
\affiliation{\MIT}
\newcommand*{\UMASS}{University of Massachusetts, Amherst, Massachusetts  01003}
\affiliation{\UMASS}
\newcommand*{\MOSCOW}{Moscow State University, General Nuclear Physics Institute, 119899 Moscow, Russia}
\affiliation{\MOSCOW}
\newcommand*{\UNH}{University of New Hampshire, Durham, New Hampshire 03824-3568}
\affiliation{\UNH}
\newcommand*{\NSU}{Norfolk State University, Norfolk, Virginia 23504}
\affiliation{\NSU}
\newcommand*{\OHIOU}{Ohio University, Athens, Ohio  45701}
\affiliation{\OHIOU}
\newcommand*{\ODU}{Old Dominion University, Norfolk, Virginia 23529}
\affiliation{\ODU}
\newcommand*{\PITT}{University of Pittsburgh, Pittsburgh, Pennsylvania 15260}
\affiliation{\PITT}
\newcommand*{\RPI}{Rensselaer Polytechnic Institute, Troy, New York 12180-3590}
\affiliation{\RPI}
\newcommand*{\RICE}{Rice University, Houston, Texas 77005-1892}
\affiliation{\RICE}
\newcommand*{\URICH}{University of Richmond, Richmond, Virginia 23173}
\affiliation{\URICH}
\newcommand*{\SCAROLINA}{University of South Carolina, Columbia, South Carolina 29208}
\affiliation{\SCAROLINA}
\newcommand*{\JLAB}{Thomas Jefferson National Accelerator Facility, Newport News, Virginia 23606}
\affiliation{\JLAB}
\newcommand*{\UNIONC}{Union College, Schenectady, NY 12308}
\affiliation{\UNIONC}
\newcommand*{\VT}{Virginia Polytechnic Institute and State University, Blacksburg, Virginia   24061-0435}
\affiliation{\VT}
\newcommand*{\VIRGINIA}{University of Virginia, Charlottesville, Virginia 22901}
\affiliation{\VIRGINIA}
\newcommand*{\WM}{College of William and Mary, Williamsburg, Virginia 23187-8795}
\affiliation{\WM}
\newcommand*{\YEREVAN}{Yerevan Physics Institute, 375036 Yerevan, Armenia}
\affiliation{\YEREVAN}
\newcommand*{\NOWCNU}{Christopher Newport University, Newport News, Virginia 23606}
\newcommand*{\NOWOHIOU}{Ohio University, Athens, Ohio  45701}
\newcommand*{\NOWUNH}{University of New Hampshire, Durham, New Hampshire 03824-3568}
\newcommand*{\NOWMOSCOW}{Moscow State University, General Nuclear Physics Institute, 119899 Moscow, Russia}
\newcommand*{\NOWSCAROLINA}{University of South Carolina, Columbia, South Carolina 29208}
\newcommand*{\NOWUMASS}{University of Massachusetts, Amherst, Massachusetts  01003}
\newcommand*{\NOWURICH}{University of Richmond, Richmond, Virginia 23173}
\newcommand*{\NOWECOSSEE}{Edinburgh University, Edinburgh EH9 3JZ, United Kingdom}
\newcommand*{\NOWGEISSEN}{Physikalisches Institut der Universitaet Giessen, 35392 Giessen, Germany}
\newcommand*{\NOWECOSSEG}{University of Glasgow, Glasgow G12 8QQ, United Kingdom}
\newcommand*{\NOWDECEASED}{deceased}

\author {Y.~Prok} 
\affiliation{\CNU}
\author {P. Bosted}
\affiliation{\JLAB}
\author {V.D.~Burkert} 
\affiliation{\JLAB}
\author {A.~Deur} 
\affiliation{\JLAB}
\author {K.V.~Dharmawardane} 
\affiliation{\ODU}
\author {G.E.~Dodge} 
\email[Contact Author \ ]{gdodge@odu.edu}
\affiliation{\ODU}
\author {K.A.~Griffioen} 
\affiliation{\WM}
\author {S.E.~Kuhn} 
\affiliation{\ODU}
\author {R.~Minehart} 
\affiliation{\VIRGINIA}
\author {G.~Adams} 
\affiliation{\RPI}
% ...
\collaboration{The CLAS Collaboration}
     \noaffiliation

\date{\today}

\begin{abstract}
The spin structure functions $g_1$ for the proton
and the deuteron have been measured over a wide kinematic range in 
$x$ and 
\Q2 using
1.6 and 5.7 GeV longitudinally polarized electrons 
incident upon polarized NH$_3$ and ND$_3$
targets at Jefferson Lab.  Scattered electrons were detected
in the CEBAF Large Acceptance Spectrometer,
for $0.05 < Q^2 < 5 $\ GeV$^2$ and $W < 3$ GeV.
The first moments of $g_1$ for the proton and deuteron
are presented -- both 
have a negative slope at low \Q2, as predicted by the extended
Gerasimov-Drell-Hearn sum rule. 
The first extraction of the generalized forward
spin polarizability of the proton $\gamma_0^p$ is also reported.
This quantity shows strong \Q2 dependence at low \Q2. 
%
Our analysis of the \Q2 evolution of the first moment of
$g_1$ shows agreement in leading order with Heavy Baryon
Chiral Perturbation Theory.
However, a significant discrepancy is observed 
between the $\gamma_0^p$ data and Chiral Perturbation
calculations for $\gamma_0^p$, even at the lowest \Q2.

\end{abstract}
\maketitle


Fundamental to our understanding of nuclear matter is a complete
picture of the spin structure of the nucleon.  
The spin of the nucleon arises from the spin 
and orbital angular momenta of both the quarks and gluons.
One way to access the quark spins in lepton
scattering is through measurements
of the spin structure functions $g_1$ and $g_2$ 
\cite{Chen05}, 
which are not well known at low
momentum transfer to the target nucleon ($Q^2 < 2$ GeV$^2$). 
At larger momentum transfer, $g_1(x,Q^2) = 
\frac{1}{2}\Sigma e_i^2\Delta q_i(x)$
(in the parton picture),
where $\Delta q_i$/$q_i$ is the net helicity of
quarks of flavor $i$ in the 
direction of the 
(longitudinally polarized) 
nucleon spin, $q_i$
is the probability of finding a quark of flavor $i$ 
with momentum fraction $x$, and $e_i$ is the quark charge.
(The Bjorken scaling 
variable $x=\frac{Q^2}{2M\nu}$ 
in the lab frame, $M$ is the nucleon mass and 
$\nu$ is the energy transferred from the electron to the target nucleon.)
At sufficiently small \Q2, $g_1$ and its moments can be more economically described
by hadronic degrees of freedom and effective low-energy approximations to QCD,
like Chiral Perturbation Theory (\ChPT).

There is particular interest in the first moment of $g_1$,
$\Gamma_1(Q^2) = \int_0^{x_0} g_1(x,Q^2)dx$, which is related to
the fraction of the nucleon spin carried by quark spins.  
The upper limit
of the integral, $x_0$, corresponds to pion production
threshold.  This limit excludes
elastic scattering, which otherwise
dominates the low \Q2 behavior of the integral.  
$\Gamma_1$ is constrained
as $Q^2 \rightarrow 0$ by the Gerasimov-Drell-Hearn (GDH) sum rule
\cite{Gerasimov,DrellHearn} to be $-\frac{\kappa^2}{8M^2}Q^2$,
where $\kappa$ is the anomalous magnetic moment
of the nucleon.
At high \Q2, $\Gamma_1$ has been measured
in deep inelastic scattering (DIS) experiments at SLAC
\cite{E143Long,E155Q2},
CERN \cite{SMCp,SMCd,COMPASSG1d} and DESY \cite{HERMES2007}.  
Ji and
Osborne \cite{JiOsborne} have
shown that the GDH sum rule can be generalized to all \Q2 via
\begin{equation}
\Gamma_1(Q^2) = {\frac{Q^2}{8}}S_1(\nu=0,Q^2) - \Gamma_1^{el}(Q^2),
\end{equation}
where $S_1(\nu,Q^2)$
is the spin-dependent virtual
photon Compton amplitude and $\Gamma_1^{el}$ is the contribution
to the integral 
from elastic scattering. 
At high $Q^2$, $S_1$ can be calculated using the operator product expansion
(OPE). By
comparing the OPE 
twist series with 
$\Gamma_1$, one can extract higher twist parameters 
\cite{DeurBjorken,twistA,twistB,SimulaDualg1,Osipenko:global},
which are sensitive to
quark-gluon and quark-quark correlations in the
nucleon at moderate \Q2.
Lattice QCD calculations may eventually be available in the moderate
\Q2 region below the range of applicability of the OPE.
At low \Q2,
$S_1$ can be calculated in \ChPT, a model-independent effective
field theory \cite{BernardReview}, 
but it is not clear how high in \Q2 these calculations
can be applied \cite{Bernard:ChPT,JiChPT}.  
Thus  $\Gamma_1$ presents a calculable 
observable that spans the
entire energy range from fundamental degrees of freedom (quarks and gluons)
to effective ones (hadrons).

Higher moments of $g_1$ are interesting as well.  
In our kinematic domain, these moments 
emphasize the resonance region over DIS kinematics
because of extra factors of $x$ in the integrand.  The
fundamental generalized forward spin polarizability of the nucleon is given by
\cite{DPV03}
\begin{equation}
\gamma_0(Q^2) = C(Q^2)\!\! \int_0^{x_0}\!\!
x^2\! \left\{ g_1(x,Q^2\!) -\! \frac{4M^2}{Q^2} x^2 g_2(x,Q^2) \right\}dx,
\end{equation}
where 
the kinematic factor $C(Q^2) = 16\alpha M^2/Q^6$ and 
$\alpha$ is the fine structure constant.
At high $Q^2$ one would expect $g_2$ to
diminish significantly 
and $g_1$ to vary logarithmically with \Q2,
thus $\gamma_0$ weighted by $Q^6$ should be largely independent
of $Q^2$ \cite{DPV03,DT04,Chen05}.  A measurement 
of $\gamma_0$ on the neutron
indicates no evidence for such ``scaling" below $Q^2$ = 1 GeV$^2$,
and furthermore the data barely agree with \ChPT\
calculations at low $Q^2$ \cite{HallAspinpol}.  
No measurement of $\gamma_0$ on the proton has been reported so far.

In order to advance our theoretical understanding of the nucleon
spin, it is essential to have data on the spin structure functions at low \Q2 and
in the resonance region, as well as at DIS kinematics.  
Data in the resonance region are necessary to calculate moments,
especially at low and moderate \Q2. 
Until recently,
data in the resonance region
were quite scarce \cite{E143res}, but  
new measurements of spin structure functions 
in the resonance region have now been reported on proton \cite{Renee,RSS},
deuteron \cite{Junho} and $^3$He targets \cite{HallAGDH1,HallAGDH2}
from Jefferson Lab.  
Testing \ChPT\ at low \Q2 
has increasingly been a focus of new spin structure experiments
\cite{EG4p,EG4d,HallALowQ2}.

The EG1 experiment of the CLAS collaboration has collected new data 
using longitudinally polarized 1.6 and 5.7 GeV electrons 
on proton (\NH3) and deuteron (\ND3) targets \cite{EG1target}.
These data cover
a wide kinematic
range that includes
invariant mass $W^2 = M^2+2M\nu-Q^2$ from 
elastic scattering (quasielastic for the deuteron)
up to 9 GeV$^2$ \cite{VipuliA1,EG1bDuality}.  
First results for the generalized forward
spin polarizability of the proton 
and
new results for the first moments of $g_1^p$ and $g_1^d$
at low and intermediate \Q2 in the range $0.05 < Q^2 < 3$
GeV$^2$ are reported in this letter.


In the EG1 experiment,
the beam was produced from a strained GaAs
wafer and had an average polarization of 70\% as measured
by Moller polarimetry \cite{Mecking:CLAS}.
The polarization of the electrons
was flipped at 30 Hz pseudo-randomly. 
The beam was rastered over the face of the
target cell to avoid heating and depolarization.  The
current varied from 0.3 nA to 10 nA depending on the beam
conditions and target.

The product of the beam and target polarizations $P_bP_t$ was 
determined from the data through comparison with the known
elastic scattering asymmetry 
and ranged from 0.50 to 0.60 
for the \NH3 target and from 0.12 to 0.23 for the
\ND3 target.
Data were also taken with $^{12}$C, $^4$He
and frozen $^{15}$N to determine the dilution
from unpolarized materials \cite{PeterRob15N}.

Scattered electrons were detected
in the CEBAF Large Acceptance Spectrometer (CLAS) in
Hall B \cite{Mecking:CLAS}, covering a range in polar angle 
from 8$^{\circ}$ to 45$^{\circ}$ in the efficient region of the detector.
Data acquisition was triggered by
a coincidence between the \Cerenkov\ detector and the calorimeter in any one
of the six sectors.
Only electrons detected in a region of
the \Cerenkov\ detector with an
efficiency greater than 80\%
were used in the analysis.
Additional details about the experiment can be found in
Refs. \cite{VipuliA1,Junho}.


We measured the raw inclusive double spin asymmetry
with longitudinally polarized beam and target
in each \Q2 and $W$ bin.  This raw asymmetry 
was then corrected for the 
difference in accumulated charge in the two beam polarization states,
$e^+e^-$ pair production and pion contamination.  
Polarization and dilution factors were divided out
and radiative corrections applied. 
The resulting asymmetry, $A_{//}$, is proportional to a linear 
combination of the two virtual
photon asymmetries $A_1$ and $A_2$ \cite{Junho}. 
Using a parameterization
of the world data 
to model $A_2$ \cite{Junho}, the unpolarized 
structure function $F_1$ \cite{ChristyBostedp,ChristyBostedd}, and 
the ratio of transverse to longitudinal
structure functions $R$ \cite{ChristyBostedp}, $A_1$  
and $g_1$ were extracted using:
\begin{align}
A_1(x,Q^2) &= \frac{1}{D}A_{//} - \eta A_2, 
\\ 
g_1(x,Q^2) &= \frac{F_1}{1+\gamma^2}(A_1 + \gamma A_2),
\label{eq:g1}
\end{align}
where the depolarization factor $D$ depends on $R$, $\eta$ 
is a kinematical factor and $\gamma^2 = \frac{Q^2}{\nu^2}$. 
The generalized forward spin polarizability for the proton was
calculated from the data for $A_1$ and the $F_1$ parameterization
using $\gamma_0(Q^2) = C\int_0^{x_0} A_1 F_1 x^2 dx$, which is
equivalent to Eq. (2).


The total systematic error on $g_1$ varies greatly depending on the kinematic bin; for
the proton it is roughly 10\% and for the deuteron it is typically
15\% for the 1.6 GeV data and 20\% for the high energy data.
The systematic error is dominated
by model uncertainties on $A_2$, $F_1$ and $R$, 
which are estimated by using different
parametrizations of the world data.  
For the deuteron data
the uncertainty in $P_bP_t$ also contributes substantially to the
systematic error.



\begin{figure}[ht]
\includegraphics[height=3.5in]{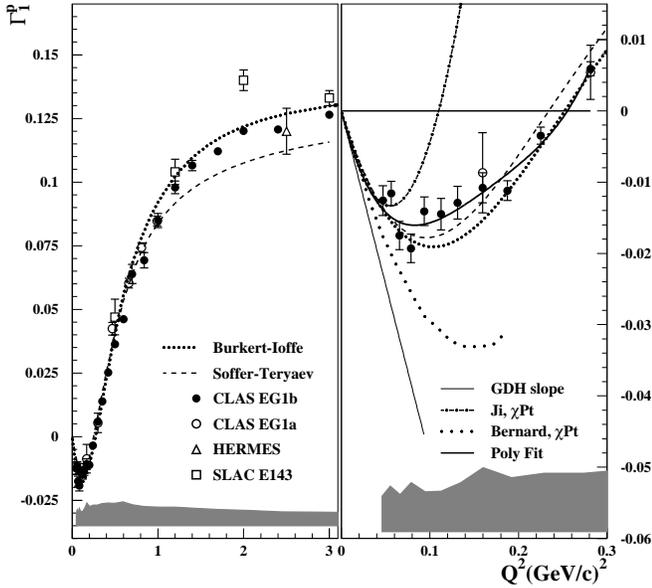}
\caption{\label{gamma1p}
$\Gamma_1^p$ as a function of
$Q^2$.  The data reported here (EG1b) are 
shown as the solid cirlces, along with the
earlier EG1 data (EG1a)\cite{Renee}, SLAC \cite{E143res} and
Hermes data \cite{HERMES2007},
shown for comparison.
The filled
circles represent the present data, including an
extrapolation over the unmeasured part of the
$x$ spectrum using a model of world data.
Phenomenological models of Burkert and Ioffe 
\cite{BurkertIoffe,BurkertIoffe2}
and Soffer and Teryaev \cite{Soffer2004}
are represented by solid and dashed lines, respectively.
The grey band represents the systematic error.
In the right plot, the scales are expanded
and $\chi$PT calculations from Bernard \cite{Bernard:ChPT} and
Ji \cite{JiChPT} are included. }
\end{figure}

\begin{figure}[ht]
\includegraphics[height=3.5in]{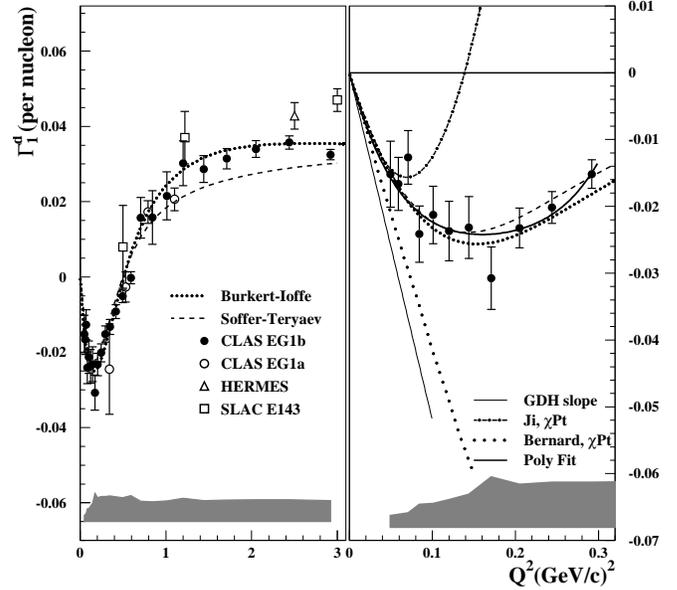}
\caption{\label{gamma1d} $\Gamma_1^d/2$ as a function of \Q2.  The
symbols and curves are the same as for 
Fig.~\ref{gamma1p}.}
\end{figure}

The values of $g_1^p$ and $g_1^d$ were extracted for \Q2 from
0.05 to 5 GeV$^2$ and for $x$ greater than 0.1;  
all results are available from the CLAS database \cite{CLASdb}.
At low \Q2, the $\Delta(1232)$ resonance is quite prominent, 
with a negative asymmetry as expected for this transition. 
It decreases steadily in strength as $Q^2$ increases.
In the mass region above the $\Delta(1232)$ resonance,
$g_1$ increases from nearly zero to large positive values as \Q2 increases.
In the $\Delta(1232)$ region and at low $Q^2$, $g_1^d/2$ is
consistent with $g_1^p$, 
as expected for a transition to an isospin $\frac{3}{2}$ state.
However, at
high \Q2, $g_1^p$ is significantly larger than
$g_1^d/2$, indicating a negative contribution from the neutron.

The first moments of $g_1^p$ and $g_1^d$ are shown in 
Figs.~\ref{gamma1p} and \ref{gamma1d}, respectively.
The parametrization of world data \cite{CLASdb} 
is used to include the
unmeasured contribution to the integral
down to $x$ = 0.001.  
The systematic uncertainty (shown by the grey bands) includes the model uncertainty
from the extrapolation to the unmeasured region.
Only the \Q2 bins in which the measured part 
(summed absolute value of the integrand)
constitutes at least 50\% of the total integral are shown. 
For the proton, the  parametrization is also used
at high $x$ (in the range $1.09 < W <$ 1.14 (1.15) GeV
for the 1.6 (5.7) GeV data). 
For the deuteron, the integration is 
carried out up to the nucleon pion production
threshold at high $x$, excluding the quasi-elastic and 
electro-disintegration contributions.
Our low \Q2 coverage
allows us to observe, for the first time, the 
slope changing sign at low $Q^2$, consistent with the expectation of a negative
slope given by the GDH sum rule at very low \Q2.
In general the data are well described by the 
phenomenological models of
Burkert and Ioffe \cite{BurkertIoffe,BurkertIoffe2} 
and Soffer and Teryaev \cite{Soffer2004}.

The low \Q2 $\Gamma_1$ data are shown in more detail
in the right-hand panels of Figs.~\ref{gamma1p} and \ref{gamma1d}.  
It is possible to make a quantitative comparison between our results
for $\Gamma_1^p$ and $\Gamma_1^d$ at low $Q^2$ and the next-to-leading
order \ChPT\ calculation by Ji, Kao and
Osborne \cite{JiChPT}, who find
$\Gamma_1^p(Q^2) = -\frac{\kappa_p^2}{8M^2}Q^2 + 3.89Q^4 + \cdots$ and 
$\Gamma_1^n(Q^2) = -\frac{\kappa_n^2}{8M^2}Q^2 + 3.15Q^4 + \cdots$.
Treating the deuteron as the incoherent sum of a proton and a neutron,
and correcting for the $D$-state as discussed in Ref.~\cite{Ciofi},
\begin{equation}
\Gamma_1^d(Q^2) = \frac{1}{2}(1 - 1.5\omega_D) \left\{ \Gamma_1^p(Q^2) 
+ \Gamma_1^n(Q^2) \right \},
\end{equation} 
where $\omega_D = 0.056$ is the weight of the $D$-wave in the deuteron,
one finds that $\Gamma_1^d(Q^2) = -0.451Q^2 + 3.26Q^4$.
In the range of $Q^2$ from 0 to 0.3 GeV$^2$, we fit
$\Gamma_1^p$ and $\Gamma_1^d$ 
to a function of the form
$aQ^2 + bQ^4 + cQ^6 + dQ^8$,
where $a$ is fixed at 
$-0.456$ (proton) and $-0.451$ (deuteron) by the GDH sum rule.  
Note that the GDH sum rule on the deuteron 
here excludes the two-body breakup part, which
otherwise nearly cancels the inelastic contribution
\cite{ArenhovelDeut}.
The fit results for the proton,
$b = 4.31 \pm 0.31$ (stat) $\pm 1.36 $ (syst), and for the
deuteron, $b = 3.19 \pm 0.44$ (stat) $\pm 0.68$ (syst), 
are both 
consistent with the $Q^4$ term predicted by Ji \etal{}
\cite{JiChPT}.  
The fit (labelled ``Poly Fit") is shown in the 
right-hand panels of Figs.~\ref{gamma1p} and
\ref{gamma1d} along with Ji's prediction. Clearly the 
$Q^6$ term becomes important even below $Q^2$ = 0.1
GeV$^2$ and this term needs to be included in the \ChPT\
calculations in order to extend the range of their validity
beyond roughly \Q2 = 0.06 GeV$^2$. 
The \ChPT\ 4$^{th}$ order (one-loop), relativistic calculation by 
Bernard \etal\ \cite{Bernard:ChPT}
is also shown in Figs.~\ref{gamma1p} and \ref{gamma1d}. 
Not shown is the result from Bernard \etal\ that includes
an estimate of the $\Delta(1232)$ and 
vector meson degrees of freedom, which are
important at low $Q^2$.  That result has large
uncertainties 
and is consistent with our data.

\begin{figure}[ht]
\includegraphics[height=3.5in]{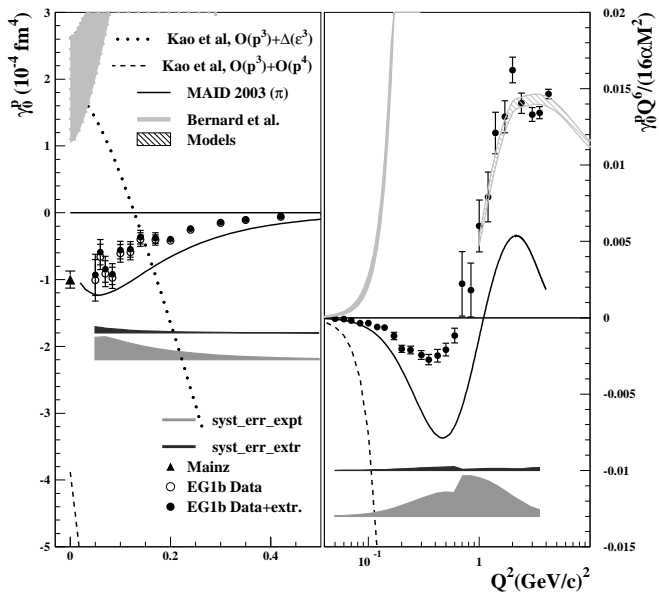}
\caption{\label{gamma0p} Generalized forward spin polarizability $\gamma_0^p$ 
as a function of \Q2 for the full integral
(closed circles), 
the measured portion of the integral (open circles) and $Q^2 = 0$
\cite{Mainzgamma0}
(triangle).  The systematic error on the measured (grey) and
unmeasured (dark) contributions are indicated by bands. 
\ChPT\ calculations 
\cite{Bernard:ChPT,ChPTKao}
are shown along with
MAID 2003 \cite{MAID2003}. 
The data shown on the right are 
weighted by $Q^6/(16\alpha M^2)$.
Our parametrization of world data 
is also shown at moderate to high \Q2.
}
\end{figure}

Fig.~\ref{gamma0p} shows the result for the generalized
forward
spin polarizability of the proton $\gamma_0^p(Q^2)$.  
Since $\gamma_0$ is weighted by an additional
factor of $x^2$ compared to $\Gamma_1$, the integral
is mostly saturated by the $\Delta(1232)$ resonance and uncertainties due
to the low-$x$ extrapolation are greatly reduced.
The MAID 2003 \cite{MAID2003} model follows the trend of the data
but lies systematically below them.
The MAID model
is consistent with our data for $A_1$ in the $\Delta$ resonance
region, 
but MAID includes only single-pion production channels, 
which leads to an underestimation of the unpolarized structure 
function $F_1$ entering the definition of $\gamma_0$.

Unlike $\Gamma_1$, $\gamma_0$ is not constrained
at $Q^2 = 0$
and is therefore a more stringent test of 
Chiral Perturbation calculations.
The leading order heavy baryon \ChPT\ calculation
by Kao, Spitzenberg
and Vanderhaeghen \cite{ChPTKao}, shown by the dotted line in
Fig.~\ref{gamma0p}, includes the $\Delta$ resonance contribution.
Their 4$^{th}$ order calculation (dashed line)
is of opposite sign
and shows no sign of convergence; neither calculation reproduces the trend or 
magnitude of the data.
The relativistic \ChPT\ calculation 
of Bernard, Hemmert and Meissner converges better at 4$^{th}$ order
\cite{Bernard:ChPT}.  That calculation, including the resonance contribution,
is represented by the grey band in Fig.~\ref{gamma0p}, 
and is also in serious disagreement with the data.
The $\Delta(1232)$ and vector meson contribution is negative
(around $-2 \times 10^{-4}$ fm$^4$) and is consistent with the
calculation by Kao \etal{} at \Q2 = 0, suggesting that the discrepancy
at low \Q2 is mainly due to the non-resonance terms \cite{MVprivate}.

In the right-hand panel of Fig.~\ref{gamma0p}, $\gamma_0^p$
is weighted by a factor of $Q^6/(16\alpha M^2)$.
In the limit of very large \Q2, this expression converges 
to the third moment of $g_1$, $a_2$, which is expected to 
scale approximately in the framework of OPE.
Our data seem to be leveling off above $Q^2 = 1.5$ GeV$^2$, 
but do not go high enough in \Q2 to confirm scaling behavior.  
We also show an evaluation
of $\int_0^{x_0} A_1 F_1 x^2 dx$ (shaded band) based on our 
model for $F_1$ and a fit to the world data for $A_1$ 
(mostly from SLAC, SMC, and HERMES in addition to our own data).  
The width of the 
shaded band indicates the combined one-sigma uncertainty 
of the models of $F_1$ and $A_1$.  Our model confirms the
leveling-off around $Q^2 = 2$ GeV$^2$ and shows a 
logarithmic fall-off at higher \Q2.


In summary, $g_1(x,Q^2)$ for the proton and the deuteron
have been measured
over a vastly expanded kinematic range at low
and intermediate momentum transfer, 
which includes the entire resonance region 
and part of the DIS regime.  
These measurements enable us to evaluate moments of $g_1$ over
a wider range in \Q2, decreasing extrapolation uncertainties. 
The first extraction of $\gamma_0^p$ has been reported along with a
new precise mapping of $\Gamma_1^p$ and $\Gamma_1^d$ down to lower
\Q2 than previously available.  At moderately high \Q2 our data for
$Q^6\gamma_0^p$ seem to level off, in agreement with models and QCD
expectations, and we see the expected trend 
toward DIS results in $\Gamma_1$.
It will be interesting to extend these measurements to higher \Q2 once the 
upgraded beam energy is available at Jefferson Lab.
At low \Q2, the first moments of $g_1^p$ and $g_1^d$ exhibit a
change in the sign of the slope, to match the negative slope constraint
from the generalized GDH sum rule, 
and are consistent with \ChPT\ calculations 
for momentum transfer values up to about 0.06 GeV$^2$.
It is important to note, however, that these \ChPT\ calculations also
assume the validity of the GDH sum rule; a more sensitive test of \ChPT\
calculations is $\gamma_0(Q^2)$.  We observe that \ChPT\
calculations fail to describe our results for 
$\gamma_0^p$, even for \Q2 as low as 0.05 GeV$^2$. 
The \ChPT\ calculations are increasingly being used to extract results
from lattice QCD and it is critical to understand their range
of applicability \cite{BernardReview}.
Data for the isoscalar quantity $\gamma_0^p - \gamma_0^n$ have also been
published by our collaboration and may give additional guidance to future
theoretical work in this area \cite{DeurIsoscalar}.  
We also look forward to results from new 
experiments at Jefferson Lab, in which spin structure functions down to
$Q^2 = 0.01$ GeV$^2$ will provide a more stringent test of \ChPT\
\cite{EG4p,EG4d,HallALowQ2}.

\begin{acknowledgments}
We would like to acknowledge the outstanding efforts of the staff
of the Accelerator and the Physics Divisions at Jefferson Lab that made
this experiment possible.  This work was supported in part by 
the U.S. Department of Energy and National 
Science Foundation, 
the Italian Istituto Nazionale di Fisica Nucleare, 
the French Centre National de la Recherche Scientifique, 
the French Commissariat \`{a} 
l'Energie Atomique
and the Korean Science and Engineering Foundation.
Jefferson Science Associates operates
the Thomas Jefferson National Accelerator Facility for the
United States Department of Energy under contract DE-AC05-84ER-40150.
We would also like to thank M. Vanderhaeghen for helpful discussions.
\end{acknowledgments}

%
%



\end{document}